\documentclass[12pt]{article}
\usepackage[latin9]{inputenc}
\usepackage{geometry}
\usepackage{float}
\usepackage{mathtools}
\usepackage{amsmath}
\usepackage{amssymb}
\usepackage{graphicx}
\usepackage{setspace}
\usepackage{esint}
\usepackage[authoryear]{natbib}
\usepackage{appendix}
\input{undertilde}
\usepackage{bibunits}
\usepackage{titlesec}
\usepackage{algorithm,algorithmicx,algpseudocode,epsfig}
\usepackage{xr}

\makeatletter

\providecommand{\tabularnewline}{\\}
\floatstyle{ruled}

\usepackage{latexsym}
\usepackage{amsthm}\usepackage{amsfonts}\usepackage{graphicx}\usepackage{epsfig}
\usepackage{bm}
\newcommand*{\patchAmsMathEnvironmentForLineno}[1]{%
      \expandafter\let\csname old#1\expandafter\endcsname\csname #1\endcsname
      \expandafter\let\csname oldend#1\expandafter\endcsname\csname end#1\endcsname
      \renewenvironment{#1}%
         {\linenomath\csname old#1\endcsname}%
         {\csname oldend#1\endcsname\endlinenomath}}%
    \newcommand*{\patchBothAmsMathEnvironmentsForLineno}[1]{%
      \patchAmsMathEnvironmentForLineno{#1}%
      \patchAmsMathEnvironmentForLineno{#1*}}%
    \AtBeginDocument{%
    \patchBothAmsMathEnvironmentsForLineno{equation}%
    \patchBothAmsMathEnvironmentsForLineno{align}%
    \patchBothAmsMathEnvironmentsForLineno{flalign}%
    \patchBothAmsMathEnvironmentsForLineno{alignat}%
    \patchBothAmsMathEnvironmentsForLineno{gather}%
    \patchBothAmsMathEnvironmentsForLineno{multline}%
    }
\usepackage[mathlines,displaymath]{lineno}

\include{def}
\def\dispmuskip{\thinmuskip= 3mu plus 0mu minus 2mu \medmuskip=  4mu plus 2mu minus 2mu \thickmuskip=5mu plus 5mu minus 2mu}
\def\textmuskip{\thinmuskip= 0mu                    \medmuskip=  1mu plus 1mu minus 1mu \thickmuskip=2mu plus 3mu minus 1mu}
\def\beq{\dispmuskip\begin{equation}}    \def\eeq{\end{equation}\textmuskip}
\def\beqn{\dispmuskip\begin{displaymath}}\def\eeqn{\end{displaymath}\textmuskip}
\def\bea{\dispmuskip\begin{eqnarray}}    \def\eea{\end{eqnarray}\textmuskip}
\def\bean{\dispmuskip\begin{eqnarray*}}  \def\eean{\end{eqnarray*}\textmuskip}

\newcommand{\wh}{\widehat}

\newcommand{\ov}{\overline}

\def\E{{\mathbb E}}                         
\def\V{{\mathbb V}}

\def\d{\delta}

\renewcommand{\d}{{\rm d}}

\def\utilde#1{\mathord{\vtop{\ialign{##\crcr
$\hfil\displaystyle{#1}\hfil$\crcr\noalign{\kern1.5pt\nointerlineskip}
$\hfil\tilde{}\hfil$\crcr\noalign{\kern1.5pt}}}}}
\def\undertilde#1{\mathord{\vtop{\ialign{##\crcr
$\hfil\displaystyle{#1}\hfil$\crcr\noalign{\kern1.5pt\nointerlineskip}
$\hfil\tilde{}\hfil$\crcr\noalign{\kern1.5pt}}}}}

\usepackage[mathlines,displaymath]{lineno}

\makeatother

\begin{document}

\title{Computationally Efficient Bayesian Estimation of High Dimensional
Archimedian Copulas with Discrete and Mixed Margins }
\author{
D. Gunawan\thanks{UNSW Business School, University of New South Wales}
\and M.-N. Tran\thanks{The University of Sydney Business School}
\and K. Suzuki\thanks{School of Mathematics and Statistics, University of New South Wales}
\and J. Dick\footnotemark[3]
\and R. Kohn\footnotemark[1]
\thanks{The research of D. Gunawan and R. Kohn
was partially supported by Australian Research Discovery Grant DP150104630 and Australian Center of Excellence Grant CE140100049.	
The research of J. Dick and K. Suzuki was partially supported by Australian Research Discovery Grant DP150101770.
}
}
\maketitle
\begin{abstract}
Estimating copulas with discrete marginal distributions is challenging,
especially in high dimensions, because computing the likelihood
contribution of each observation requires evaluating $2^{J}$
terms, with $J$ the number of discrete variables. Our article focuses on the estimation of Archimedian copulas, for example, Clayton and Gumbel copulas. Currently, data augmentation methods
are used to carry out inference for discrete copulas and, in practice, the computation
becomes infeasible when $J$ is large. Our article proposes two
new fast Bayesian approaches for estimating high dimensional Archimedian copulas with
discrete margins, or a combination of discrete and continuous margins.
Both methods are based on recent advances in Bayesian methodology that work
with an unbiased estimate of the likelihood rather than the likelihood itself, and
our key observation is that we can estimate the likelihood of a discrete Archimedian copula
unbiasedly with much less computation than evaluating the likelihood exactly or with
current simulation methods that are based on augmenting the model with latent variables.
The first approach builds on the pseudo marginal method
that allows Markov chain Monte Carlo simulation from the posterior
distribution using only an unbiased estimate
of the likelihood. The second approach is based on a Variational Bayes approximation to the posterior
and also uses an unbiased estimate of the likelihood.
We show that the two new approaches enable us to carry out Bayesian inference for high values of $J$ for the Archimedian copulas where the computation was previously too expensive.
The methodology is illustrated through several
real and simulated data examples.

Key words:  Markov chain Monte Carlo; Correlated pseudo marginal Metropolis-Hastings; Variational Bayes; Archimedian copula{\small \par}

\end{abstract}

\section{Introduction}\label{S: introduction}

Copula models provide a flexible approach for modeling multivariate
distributions by capturing the joint dependence structure by a copula
and modeling the marginal distributions of the variables separately and flexibly
\citep[see, for example,][]{Trivedi2005, Smith2012}. There are now a number of copula
models that allow for a wide range of dependence. 

In many applications in the literature, multivariate data are modeled as
parametric copulas and unknown copula parameters are often estimated by maximum likelihood.
However, for high dimensional data with discrete variables, maximum
likelihood estimation (MLE) is expensive as it requires $2^{J}$ evaluations of a $J$ dimensional
cumulative distribution function to compute the probability mass function (pmf) at a single data point.
Recently,  Bayesian methods have been developed which offer, to some
extent, solutions to this problem, in particular for Gaussian copula model. \citet{Pitt2006} propose an efficient
Bayesian data augmentation method to estimate the parameters of a Gaussian copula model
with all its margins discrete. They introduce latent variables to the model
and generate these latent variables within an MCMC scheme. \citet{Murray2013} use data augmentation together with a parameter expansion approach to estimate Gaussian copula factor model with discrete margins, or a combination of discrete and continuous margins. Recently, \citet{Pakman2014} proposed an exact Hamiltonian Monte Carlo approach to sample from truncated multivariate Gaussian. This exact Hamiltonian Monte Carlo approach can be used to sample the latent variables together with parameter expanded Gibbs sampling of \citet{Murray2013} to obtain an efficient algorithm for the Gaussian copula for discrete margins, or a combination of discrete and continuous margins. 

There is much less literature on how to estimate other copula models for discrete data, for example the Archimedian copulas (see \citet{Hofert2008,Hofert2012}). \citet{Smith2012} extend the data augmentation approach to the (discrete)
D-vine copula, which is constructed from a sequence of bivariate ``pair-copulas''. They consider
Archimedean and elliptical copulas as the building blocks of the (discrete) D-vine
copula. They also extend the method to combinations of discrete and continuous marginals and give a two dimensional example. There are currently some issues with the existing Bayesian data augmentation methods for high dimensional Archimedian copulas. 
As the number of latent variables is of the same size as
the data, i.e. a matrix of size $n\times J$, with $n$ being the number
of observations and $J$ the number of dimensions, these methods 
suffer from computational issues when either $n$ or $J$
is large. This is because generating these latent variables for the Archimedean copulas is very expensive since the inverses
of the conditional distributions for the Archimedian copula model are usually unavailable in closed
form and need to be computed numerically. Furthermore, for the Archimedean copula,
the conditional copula distribution functions and their
densities are  also expensive
to compute for large $J$.
Another problem with data augmentation approaches is that for large $J$ they are likely to induce
high correlations in the MCMC iterates because the copula parameter is generated conditional on the latent variable using the Metropolis within Gibbs step. This is very inefficient if the latent variables are highly correlated with the copula parameter; see Section~\ref{S:PM and VBIL vs data augmentation}.

Our article introduces several methodological innovations for Bayesian inference in high dimensional discrete and mixed marginal for the Archimedian copulas to overcome the problems
experienced when using the latent variable data augmentation approach. 
We note, however, that our methods can be applied to any parametric copulas including Gaussian copulas.  
Our key observation is that the likelihood of a copula model is a product of
terms each of which is expensive to evaluate, but it is relatively cheap to estimate each term, and hence the likelihood, unbiasedly.
Based on this insight, we adapt to the discrete and mixed margin Archimedian copulas, two recent approaches to Bayesian inference which
work with unbiased estimates of the likelihood.
The first approach is based on the pseudo marginal (PM) method of 
 \cite{Andrieu:2009} and the second (approximate) approach is based on the Variational
Bayes with intractable likelihood (VBIL) method of \cite{Tran:2015}.  Section~\ref{S: estimation and inference} discusses these approaches.

In particular, our first contribution is to
introduce into the copula literature the pseudo marginal (PM) approaches.
These approaches include the standard PM and the correlated PM approaches discussed in section~\ref{SS: correlated pseudo} and the block
sampling method discussed in section~\ref{SS: block PM method}. 
Second, we introduce into the copula literature a variational Bayes approach that works with an unbiased estimate of the likelihood
and is much faster than the PM approaches. Although this approach is approximate, we show in our applications that the approximations are very accurate. There are other alternatives of variational inference method that can be used, in particular, the so called reparameterization trick of \citet{Kingma2014}. However, the reparameterization trick requires unbiased estimates of the gradient of the log-likelihood instead of the unbiased estimate of the likelihood. It is in general even more difficult to obtain an  accurate estimate of gradient of the log-likelihood than it is to obtain accurate estimate of the likelihood (\citet{Sherlock2015}).

The attraction of the proposed approaches is that: (i)~they can be used
for high dimensional problems (large $J$) and for large data sets
(large $n$), where computation based on latent variable methods is prohibitively expensive or infeasible;
see Section~\ref{S:simulation studies}.
Our article considers 50 dimensional real discrete data examples and simulated data examples with up to 100 dimensions.
To the best of our knowledge, the highest dimension handled in the literature is less than 30; see, for example,
16-dimension in \citet{Smith2012}), 20-dimension in \citet{Panagiotelis:2012}, and
6-dimension in \citet{Panagiotelis}.
(ii)~As we show in Section ~\ref{S:simulation studies} that the PM approaches are also much more efficient than data augmentation because they generate the copula parameter with the latent variables integrated out.

An online supplement to our article gives further technical and empirical results.
All equations, lemmas, tables, etc in the article are referred to as equation (1), lemma 1, table 1, etc, and
in the supplement they are referred to as equation (S1), lemma S1 and table S1, etc.

\section{The Copula Model }\label{S: copula model}
\subsection{Definition}\label{SS: definition}
Let $\bm X=(X_1,...,X_J)^\top$ be a vector of $J$ random variables, and $F(\bm x)$ with $\bm x = (x_1,...,x_J)^\top$ be the joint cumulative distribution function (cdf) of $\bm X$ with marginal cdf $F_j(x_j)$, $j=1,...,J$. 
We are interested in modeling $F(\bm x)$. 
A copula $C(\bm u) $ of dimension $J$ is a joint cdf defined on
$\left[0,1\right]^{J}$, that has each of its margins uniformly distributed on $\left[0,1\right]$.
In copula modeling \citep{Sklar1959}, the joint cdf $F(\bm x)$ is modeled as
\begin{align}\label{eq:joint distributions}
F\left(\bm x \right) & =  C\left(F_{1}\left(x_{1}\right),F_{2}\left(x_{2}\right),...,F_{J}\left(x_{J}\right)\right).
\end{align}
Assume that $C(\cdot)$ has the density $c(\cdot)$. If the $X_j$ are continuous, then
\begin{align}\label{eq:joint prob copula}
\Pr(\bm X \in \prod_{j=1}^J (a^X_j,b^X_j]) & = {\displaystyle \int_{a_1}^{b_1} \cdots \int_{a_J}^{b_J}} c(\bm u ) \d \bm u
\end{align}
where $a_j:=F_j(a^X_j)$ and $b_j: = F_j(b^X_j)$, for $j=1,\dots, J$.
If the $X_j$ are discrete random variables, then
\begin{align}\label{eq:joint prob copula discrete}
\Pr&(\bm X = \bm x)  = {\displaystyle \int_{a_1}^{b_1} \cdots \int_{a_J}^{b_J}} c(\bm u ) \d \bm u \nonumber \\
& = \bigg ( \prod_{j=1}^J (b_j-a_j)\bigg )  \int_{0}^{1} \cdots \int_{0}^{1} c\bigg((b_1-a_1)v_1 + a_1,\dots,(b_J-a_J)v_J + a_J\bigg) \, \d\bm v
\end{align}
with $u_j = (b_j-a_j)v_j + a_j$, $\bm v:= (v_1, \dots, v_J)^\top$ and $b_j=F_j(x_j)$, $a_j=F_j(x_j^{-})$. See, e.g., \cite{Smith2012}.
We can now apply Monte Carlo (MC) to estimate the integral unbiasedly.

We can simplify the integrals \eqref{eq:joint prob copula} and \eqref{eq:joint prob copula discrete} when some of the $a_j$ are 0,
which can be useful in terms of MC simulation as the dimension of the integral is reduced.
Without loss of generality, suppose that
$a_1, \cdots, a_K \neq 0$ and $a_{K+1}, \dots, a_{J} = 0$.
Then,
\begin{align}  \label{eq: shorter integral}
\int_{a_1}^{b_1} \cdots \int_{a_J}^{b_J} c(\bm u) \, \d\bm u
&= \int_{a_1}^{b_1} \cdots \int_{a_K}^{b_K} D(\bm u_{1:K},\bm b_{K+1:J}) \, \d\bm u_{1:K},
\end{align}
where $\bm u_{1:K} := (u_1, \dots, u_{K})$, $\bm b_{K+1:J} := (b_{K+1}, \dots, b_{J})$ and
\begin{align} \label{eq: def of D}
D(\bm u_{1:K},\bm b_{K+1:J})& := \partial_{u_1} \cdots \partial_{u_K} C(\bm u_{1:K}, \bm b_{K+1:J}):= \frac{\partial^K C(\bm u_{1:K}, \bm b_{K+1:J})}{\partial u_1 \cdots \partial u_K} .
\end{align}
We can rewrite the integral \eqref{eq: shorter integral} as
\begin{align}\label{eq: integral short in z}
& \int_{a_1}^{b_1} \cdots \int_{a_K}^{b_K}  D(\bm u_{1:K},\bm b_{K+1:J}) \, \d\bm u_{1:K} = \bigg ( \prod_{j=1}^K(b_j-a_j)\bigg ) \nonumber \\
 & \times \int_{0}^{1} \cdots \int_{0}^{1} D((b_1-a_1)v_1 + a_1,\dots,(b_K-a_K)v_K + a_K, \bm b_{K+1:J}) \, d\bm v_{1:K}
\end{align}
with $u_j = (b_j-a_j)v_j + a_j$, $j=1,...,K$,
and we can now estimate it unbiasedly using MC.
This leads to faster and more stable MC estimation as long as we can evaluate $D(\bm u_{1:K},\bm b_{K+1:J})$. 

\subsection{Examples} \label{ss: Clayton copula example}
This section gives some details on the Clayton copula that we consider in this paper. 
See Section~\ref{SS: defining Gumbel copula} for the Gumbel copula case.
The Clayton copula is 
\begin{align}\label{eq: clayton cdf}
C(\bm u):= \left( \sum_{j=1}^J u_j^{-\theta} -J +1 \right)^{-\frac{1}{\theta}}, \quad \theta>0,
\end{align}
and its density is
\begin{align} \label{eq: clayton pdf}
c(\bm u) &=  \partial_{u_1} \cdots \partial_{u_J} C(\bm u) = \prod_{k=0}^{J-1} (\theta k +1)
	\left( \prod_{j=1}^J u_j \right)^{-(1+\theta)}
	\left( \sum_{j=1}^J u_j^{-\theta} -J +1 \right)^{-(J+\frac{1}{\theta})}.
\end{align}
We use \eqref{eq: shorter integral} to evaluate the integral \eqref{eq:joint prob copula discrete} if $a_j=0$, for $j=K+1,\dots, J$ and $a_j > 0 , j=1, \dots, K$, in \eqref{eq:joint prob copula discrete}.  It is readily checked that
\[D(\bm u_{1:K},\bm b_{K+1:J})=\prod_{k=0}^{K-1} (\theta k +1)
	\left( \prod_{j=1}^{K} u_j \right)^{-(1+\theta)} \left( \sum_{j=1}^{K} u_j^{-\theta} + \sum_{j=K+1}^{J} b_j^{-\theta} -J +1 \right)^{-(K+\frac{1}{\theta})}.\]
This integration \eqref{eq: shorter integral} is preferable since $D(\bm u_{1:K},b_{K+1:J})$ is bounded on the domain of integration and the dimension of the integration is reduced.

\subsection{Mixed continuous and discrete marginals}  \label{ss: discrete and continuous marginals}

We now extend the copula framework to accommodate the case where $\boldsymbol{X}$
has both discrete and continuous marginals, with the  distribution
of $\bm X$ generated by the copula $C(\cdot)$ with density $c(\cdot)$.
Without loss of generality, suppose that $X_{1},...,X_{r}$ are the discrete marginals and $X_{r+1},...,X_{J}$ are the
continuous marginals with cdf $F_j(x_j)$ and pdf $f_j(x_j)$.  Then, similarly to \eqref{eq:joint prob copula}
\begin{align}\label{eq: mixed case general}
\Pr(\bm X_{1:r}=\bm x_{1:r}|\bm x_{r+1:J})p(\bm x_{r+1:J}) = \int_{a_1}^{b_1} \cdots \int_{a_r}^{b_r} c(u_1,...,u_r,u_{r+1},...,u_J) \d \bm u_{1:r} \prod_{j=r+1}^J f_j(x_j)
\end{align}
where $u_j=F_j(x_j)$ for $j=r+1,\cdots, J$.

\section{ Bayesian inference} \label{S: estimation and inference}
This section discusses Bayesian estimation and inference using the PM and VBIL methods.
In the statistical literature, \citet{Beaumont:2003} was the  first to propose the
PM approach, and \citet{Andrieu:2009} studied some of its theoretical
properties. The PM methods
carry out Markov chain Monte carlo (MCMC)  on an expanded space and use an unbiased estimate
of the likelihood, instead of the likelihood.
\citet{Pitt:2012} and
\citet{Doucet:2015} show that
the variance of the log of the estimated likelihood should be around
1 for the optimal performance of the standard PM method (defined more precisely in section~\ref{SS: correlated pseudo}),
and that the performance of  the standard PM deteriorates exponentially as the variance of the log of
the estimated likelihood increases beyond 1. Thus, a serious drawback of the
standard PM method is that it is highly sensitive to the variability
of the log of the estimated likelihood \citep[see, e.g.,][]{Flury:2011}.
It may therefore be very computationally demanding to ensure that
the variance of the log of the estimated likelihood is around 1 for the high
dimensional discrete Archimedian copulas. As a remedy, \citet{Deligiannidis:2015} modify
the standard PM method by correlating the pseudo-random numbers used
in constructing the estimators of the likelihood at the current and
proposed values of the Markov chain.  This correlated PM  approach helps the
chain to mix well even if highly variable estimates of the likelihood
are used. Thus, the correlated PM requires far fewer computations
at every iteration than the standard PM. \citet{Tran:2016} propose an alternative to the
correlated PM approach which samples the pseudo-random numbers in blocks and show that for some problems it can be more efficient
than the correlated PM approach of  \citet{Deligiannidis:2015}.

The VBIL method, developed by \citet{Tran:2015} and described in section \ref{SS: VBIL}, provides a fast variational
approximation of the posterior distribution when the likelihood is
intractable, but can be estimated unbiasedly. \citet{Tran:2015} show
both theoretically and empirically that the VBIL method still works well
when only highly variable estimates of likelihood are available.

\subsection{Estimating the likelihood unbiasedly}\label{ss: likelihood and unbiased estimation}
This section describes how to obtain unbiased estimates of the likelihood in copula estimation,
which are required by the PM and VBIL approaches. 
Suppose that we have $n$ observations $\bm x_t, t=1, \dots, n$. Define
$L_t(\bm \theta):= \Pr( \bm x_t|\bm \theta) $, where $\bm \theta$ is the vector of parameters in the copula model,
and $\Pr( \bm x_t|\bm \theta)$ is defined as in \eqref{eq:joint prob copula discrete}.
The likelihood is
$L(\bm \theta):= \prod_{t=1}^n L_t(\bm \theta) $. We can estimate each $L_t(\bm \theta)$ unbiasedly by MC
as
\begin{align}\label{eq: unbiased estimate integral}
\wh L_t(\bm \theta)  & =\left( \prod_{j=1}^J (b_j-a_j)\right)\times \frac1M \sum_{i=1}^M  c\bigg((b_1-a_1)u_1^{(t,i)} + a_1,\dots,(b_J-a_J)u_J^{(t,i)} + a_J\bigg),
\end{align}
where the $\bm u^{(t,i)}:= (u^{(t,i)}_1, \dots, u^{(t,i)}_J)$  are uniformly distributed random numbers, $i=1,...,M$ with $M$ the number of samples.
A similar estimator can be obtained for the integral in \eqref{eq: integral short in z}. We define the likelihood estimate as
$\wh L_M(\bm \theta): = \prod_{t=1}^n \wh L_t(\bm \theta) $. 
Given that the sets $\bm u^{(t)}:= \{ \bm u^{(t,i)}, i=1, \dots, M\}$ are independent across $t$,
it is clear that $\wh L_M(\bm \theta)$ is an unbiased estimator of $L(\bm \theta)$, i.e. $\E(\wh L_M(\bm \theta))=L(\bm \theta)$.
To indicate that $\wh L_M(\bm\theta)$ also depends on the random variates $\bm u:= \{ \bm u^{(t)}, t=1, \dots, n \}$, we will sometimes write $\wh L_M(\bm \theta)$ as $\wh L_M (\bm \theta, \bm u )$.

\subsection{The Pseudo Marginal methods}  
This section discusses the PM approaches. Let $p_U\left(\bm {u}\right)$
be the density function of $\bm {u}$ and $p_\Theta (\bm \theta) $ the prior for $\bm \theta $. We define the joint density of
$\bm \theta$ and $\bm {u}$  as
\begin{align}\label{eq: target joing density}
\ov \pi\left(\bm \theta,\bm {u}\right):=\wh L_M(\bm \theta, \bm u )p_\Theta \left(\bm \theta\right)p_U (\bm {u})/ \ov L,
\end{align}
where $\ov L:= \int L(\bm \theta) p_\Theta (\bm\theta)\d\bm\theta $ is the marginal likelihood.
Clearly,
\begin{align*}
\ov \pi(\bm \theta) & = \int \pi(\bm \theta, \bm u) \d \bm u = L(\bm\theta)p_\Theta(\bm\theta) /\ov L = \pi(\bm \theta)
\end{align*}
is the posterior of $\bm \theta$, because $\int \wh L_M(\bm \theta , \bm u) p_U (\bm u) \d \bm u = L(\bm \theta)$ by unbiasedness.
Hence, we can obtain samples from the posterior density $\pi(\bm \theta) $ by sampling $\bm \theta $ and $ \bm u$ from $\ov \pi(\bm \theta, \bm u)$.

Let $q_\Theta \left(\bm \theta^{\prime}; \bm \theta\right)$ be a proposal density for $\bm \theta'$
with current state $\bm \theta$ and $q_U(\bm u'; \bm u)$ the proposal density for $\bm u'$ given $\bm u$. We assume that $q_U(\bm u'; \bm u)$ satisfies
the reversibility condition
\begin{align} \label{eq: detailed balance u}
q_U(\bm u';\bm u) p_U(\bm u) = q_U(\bm u; \bm u')p_U(\bm u'),
\end{align}
which is clearly satisfied in the standard PM where $q_U(\bm u';\bm u) = p_U(\bm u')$.
Then, we generate a proposal $\bm \theta'$
from $q_\Theta \left(\bm \theta'; \bm \theta\right)$ and $\bm {u}'$
from $q_U\left(\boldsymbol{u'}; \bm u\right)$, and accept these proposals with the acceptance probability
\begin{align}\label{eq:acceptance MH probability}
\alpha (\bm \theta,\boldsymbol{u};\bm \theta',\boldsymbol{u}{'})& :=\min \left \{
 1,\dfrac{ \wh L_M(\bm \theta^\prime, \bm u^{\prime} )p_\Theta (\bm\theta^{'}) p_U(\bm u^\prime )}
{ \wh L_M(\bm \theta, \bm u )p_\Theta (\bm\theta) p_U(\bm u )}
\dfrac{ q_\Theta (\bm \theta; \bm \theta^\prime)q_U(\bm u; \bm u^\prime) }
{ q_\Theta (\bm \theta^\prime; \bm \theta)q_U(\bm u^\prime; \bm u) }
\right \}\notag \\
& = \min \left \{
 1,\dfrac{ \wh L_M(\bm \theta^\prime, \bm u^{\prime} )p_\Theta (\bm\theta^{'}) }
{ \wh L_M(\bm \theta, \bm u )p_\Theta (\bm\theta)}
\dfrac{ q_\Theta(\bm \theta; \bm \theta^\prime) }
{ q_\Theta(\bm \theta^\prime; \bm \theta) }
\right \}
\end{align}
using \eqref{eq: detailed balance u}.

In the standard PM method, $q_U(\bm u';\bm u) = p_U(\bm u')$ so that a new set of pseudo-random numbers
$\boldsymbol{u}'$ is generated independently of $\bm u$ each time we estimate the likelihood.
The performance of the PM approach depends on the number of samples $M$
used to estimate the likelihood.
\citet{Pitt:2012} suggest selecting $M$ such that
the variance of the log of the estimated
likelihood to be around 1 to obtain an optimal trade-off between computing
time and statistical efficiency. However, in many applications such as the high dimensional copula modelling considered in this paper, it is computationally very expensive to ensure that the variance of the
log-likelihood is around 1.

\subsubsection{The correlated PM approaches\label{SS: correlated pseudo}}
The correlated PM proposed by \citet{Deligiannidis:2015}
correlates the MC random numbers, $\bm u $, used in constructing
the estimators of the likelihood at the current and proposed values
of the parameters to reduce the variance of the difference $\log \wh L_M(\bm \theta^\prime , \bm u^\prime) - \log \wh L_M(\bm \theta , \bm u)$
appearing in the MH acceptance ratio \eqref{eq:acceptance MH probability}.
This method tolerates a much larger variance of the likelihood estimator
without the MCMC chain getting stuck. The correlated PM approach is given in Algorithm \ref{alg: correlated pseudo marginal}.
It is easy to check that the reversibility condition~\eqref{eq: detailed balance u} is satisfied under this scheme in terms of $\bm z$.

\begin{algorithm}[H] \caption{Correlated PM} \label{alg: correlated pseudo marginal}
\begin{enumerate}
\item Sample $\bm \theta'$ from $q_\Theta\left(\bm \theta'; \bm \theta\right)$
\item Sample $\boldsymbol{z}^{*}\sim N\left(0,I\right)$
and set $\boldsymbol{z}'=\phi\boldsymbol{z}+\sqrt{1-\phi^{2}}\boldsymbol{z}^{*}$,
where $\phi$ is the correlation between $\boldsymbol{z}=\Phi^{-1}(\bm u)$ and $\boldsymbol{z}'$
and is set close to $1$. Set $\bm u'=\Phi\left(\boldsymbol{z}'\right)$. Here, $\Phi$ denotes the standard normal cdf. 
\item Compute the estimate $\wh L_M(\bm \theta^\prime , \bm u^\prime) $.
\item Accept the proposal $ (\bm \theta',\boldsymbol{u}')$
with probability given in \eqref{eq:acceptance MH probability}.
\end{enumerate}
\end{algorithm}

\subsubsection{The block PM approach} \label{SS: block PM method}
The block PM approach of \citet{Tran:2016} is an alternative to the correlated PM by
updating  $\boldsymbol{u}$ in blocks. Suppose that
$\boldsymbol{u}$ is partitioned into $G$ blocks $\boldsymbol{u}_{\left(1\right)},...,\boldsymbol{u}_{\left(G\right)}$. We write the
target density in $\bm \theta$ and $\bm u $ as
\begin{align}  \label{eq: target blockwise}
\pi\left(\bm \theta,\boldsymbol{u}\right)\coloneqq \wh L_M(\bm \theta, \bm u_{(1)}, \dots, \bm u_{(G)} ) p_\Theta( \bm \theta) p_U(\bm u_{(1)} , u_{(2)} , \dots,
\bm u_{(G)} )/\ov L
\end{align}
Instead of updating the full set of $\left(\bm \theta,\boldsymbol{u}\right)$
at each iteration of the PM algorithm, the block PM algorithm
updates $\bm\theta$ and a block $\boldsymbol{u}_{\left(k\right)}$
at a time.
Block PM always takes less CPU time in each MCMC iteration than the standard
and correlated PM approaches as it does not generate the entire set of random numbers $\bm u$.
The block index $k$ is selected at random from $1,...,G$
with $\Pr\left(K=k\right)>0$ for every $k=1,...,G$. Our article uses
$\Pr\left(K=k\right)=1/G$. Using this scheme, the acceptance
probability becomes
\begin{align*}
\min\left\{ 1,
\dfrac { \wh L_M(\bm \theta^\prime, \bm u_{(1)}, \dots, \bm u_{(k-1)}, \bm u_{(k)}^\prime , \bm u_{(k+1)}, \dots, \bm u_{(G)}) p_\Theta(\bm \theta) }
{ \wh L_M(\bm \theta, \bm u_{(1)}, \dots, \bm u_{(k-1)}, \bm u_{(k)} , \bm u_{(k+1)}, \dots, \bm u_{(G)}) p_\Theta(\bm \theta)  }
 \times \dfrac{ q_\Theta(\bm \theta; \bm \theta^\prime) }{ q_\Theta(\bm \theta^\prime ; \bm \theta)}
\right \}.
\end{align*}

\subsection{Variational Bayes with Intractable Likelihood (VBIL) }\label{SS: VBIL}
Variational Bayes (VB) is a fast method to approximate the posterior
distribution $\pi\left(\bm \theta\right)$ by a distribution $q_{\bm \lambda}\left(\bm \theta\right)$
within some tractable class, such as an exponential family, where
$\bm \lambda$ is a variational parameter which is chosen
 to minimise the Kullback-Leibler divergence between
$q_{\bm \lambda}\left(\theta\right)$ and $\pi\left(\bm \theta\right)$ \citep{Ormerod:2010}
\begin{align*}
KL\left(\bm \lambda\right) & =  KL\left(q_{\bm \lambda}\left(\bm \theta\right)||\pi\left(\bm \theta\right)\right)
 := \int\log\frac{q_{\bm \lambda}\left(\bm \theta\right)}{\pi\left(\bm \theta\right)}q_{\bm \lambda}\left(\bm \theta\right)\d\bm \theta.
\end{align*}
Most  current VB algorithms require that the likelihood $L(\bm \theta) $
is computed analytically for any $\bm \theta$. \citet{Tran:2015}
proposed the VBIL algorithm that works with an unbiased estimate of
the likelihood. Define $z:=\log\wh L_M(\bm \theta, \bm u) -\log L(\bm \theta)$
so that $\wh L_M(\bm \theta, \bm u) = L(\bm \theta) \exp (z),$
and denote by $g(z|\bm \theta )$ the density of $z$ given
$\bm \theta$.  The reason for introducing $z$ is that
it is easier to work with a scalar $z$ rather than the high dimensional
random numbers $\boldsymbol{u}$.
In this section we also write $\wh L_M(\bm \theta, \bm u)$ as $\wh L_M(\bm \theta, z)$.
Due to the unbiasedness of the estimator $\wh L_M(\bm \theta,\bm u  )$,
we have $\int\exp\left(z\right)g\left(z|\bm \theta\right)\d z =1$.
We now define the corresponding target joint density of $\bm \theta$ and
$z$ as
\begin{align*}
\ov \pi(\bm \theta,z):=L(\bm \theta) p_\Theta(\bm \theta) \exp (z )g (z|\bm \theta )/ \ov L =\pi (\bm \theta)\exp (z)g(z|\bm \theta)
\end{align*}
which admits the posterior
density $\pi\left(\bm \theta\right)$ as its marginal. \citet{Tran:2015}
approximate $\ov \pi\left(\bm \theta,z\right)$ by
$
q_{\bm \lambda}\left(\bm \theta,z\right):=q_{\bm \lambda}(\bm \theta)g\left(z|\bm \theta\right),
$
where $\bm \lambda$ is the vector of  variational parameters that are estimated
by minimising the Kullback-Leibler divergence between $q_{\bm \lambda}\left(\bm \theta,z\right)$
and $\ov\pi\left(\bm \theta,z\right)$
in the augmented space, i.e.,
\[
KL\left(\bm \lambda\right)=KL\left(q_{\bm \lambda}\left(\bm \theta,z\right)||\ov\pi\left(\bm \theta,z\right)\right)\coloneqq\int q_{\bm \lambda}(\bm \theta)g\left(z|\bm \theta\right)\log\dfrac{q_{\bm \lambda}(\bm \theta)g\left(z|\bm \theta\right)}{\ov\pi\left(\bm \theta,z\right)}\d z \d \bm \theta.
\]
The gradient of $KL\left(\bm \lambda\right)$
is
\begin{align} \label{eq: gradient of KL}
\nabla_{\bm \lambda}KL\left(\bm \lambda\right) & =  \E_{q_{\bm \lambda}} \bigg\{ \nabla_{\bm \lambda}\left[\log q_{\bm \lambda}\left(\bm \theta\right)\right]\left(\log q_{\bm \lambda}\left(\bm \theta\right)-\log\left(p_\Theta\left(\bm \theta\right) \wh L_M(\bm \theta, z)\right)\right)\bigg \} ,
\end{align}
where the expectation is with respect to $q_{\bm \lambda}\left(\bm \theta,z\right)$.
See \citet{Tran:2015} for details.
We  obtain an
unbiased estimator $\widehat{\nabla_{\bm \lambda}KL}\left(\bm \lambda\right)$
of the gradient $\nabla_{\bm \lambda}KL\left(\bm \lambda\right)$
by generating $\bm \theta\sim q_{\bm \lambda}\left(\bm \theta\right)$
and $z\sim g\left(z|\bm \theta\right)$ and computing the likelihood estimate
$\wh L_M(\bm \theta , z )$. MC method can be used to estimate the gradient unbiasedly
and stochastic optimization is then used to find the optimal $\bm \lambda$.

Algorithm~\ref{alg: VBIL alg} gives general pseudo code for the VBIL method. We note that each iteration of the algorithm
can be parallelized because the gradient is estimated by importance sampling.
\begin{algorithm} \caption{The VBIL algorithm} \label{alg: VBIL alg}
Initialise $\bm \lambda^{(0)}$ and let $S$
be the number of samples used to estimate the gradient \eqref{eq: gradient of KL}.
\begin{enumerate}
\item Initialisation: Set $t=0$
\begin{enumerate}
\item Generate $\bm \theta_{s}^{\left(t\right)}\sim q_{\bm \lambda}\left(\bm \theta\right)$
and $z_{s}^{\left(t\right)}\sim g\left(z|\bm \theta\right)$, for $s=1,...,S$
\item  Denote $\widehat{h}\left(\boldsymbol{\theta},z\right)=\log\left(p_{\Theta}\left(\boldsymbol{\theta}\right)\widehat{L}_{M}\left(\boldsymbol{\theta},z\right)\right)$ and set
\begin{align*}
c^{(t)} & =\frac{
\wh {\mathbb{C}\mathrm{ov}} \big(\widehat{h}\left(\boldsymbol{\theta},z\right)\nabla_{\bm \lambda}\log q_{\bm \lambda}
(\bm \theta),
\nabla_{\bm \lambda}\log q_{\bm \lambda}(\bm \theta)\big)} {\wh {\V}\big(\nabla_{\bm \lambda}\log q_{\bm \lambda}(\bm \theta)\big)},
\end{align*}
where $\wh {\mathbb{C}\mathrm{ov}} (\cdot )$ and $\wh \V\left(\cdot \right)$ are sample estimates of covariance
and variance based on the samples $\left(\bm \theta_{s}^{\left(t\right)},z_{s}^{\left(t\right)}\right)$,
for $s=1,...,S$. The control variate $c^{\left(t\right)}$ is employed
to reduce the variance in the gradient estimation.
\end{enumerate}
\item Cycle: Repeat the following until a stopping criterion is satisfied.
\begin{enumerate}
\item Set $t=t+1$ and generate $\bm \theta_{s}^{\left(t\right)}\sim q_{\bm \lambda}\left(\bm \theta\right)$
and $z_{s}^{\left(t\right)}\sim g\left(z|\bm \theta\right)$, for $s=1,...,S$
\item Estimate the gradient
\[
\nabla_{\boldsymbol{\lambda}}KL\left(\boldsymbol{\lambda}\right)^{(t)}=\frac{1}{S}\sum_{s=1}^{S}\nabla_{\boldsymbol{\lambda}}\log q_{\boldsymbol{\lambda}}\left(\boldsymbol{\theta}_{s}^{\left(t\right)}\right)\left(\log q_{\boldsymbol{\lambda}}\left(\boldsymbol{\theta}_{s}^{\left(t\right)}\right)-\widehat{h}\left(\boldsymbol{\theta}_{s}^{\left(t\right)},z_{s}^{\left(t\right)}\right)-c^{\left(t-1\right)}\right).
\]

\item Estimate the control variate $c^{\left(t\right)}$ as in step 1(b).
\item Update the variational parameter $\bm \lambda$ by
\[
{\bm \lambda}^{\left(t+1\right)}=\bm\lambda^{\left(t\right)}-a_{t}I_{F}\left(\bm\lambda^{\left(t\right)}\right)^{-1}\nabla_{\boldsymbol{\lambda}}KL\left(\boldsymbol{\lambda}\right)^{(t)}
\]

The learning rate sequence $\{a_{t}, t \geq 1, a_{t}>0\}$
satisfies the Robbins-Monro conditions $\sum_{t}a_{t}=\infty$ and
$\sum_{t}a_{t}^{2}<\infty$ \citep{Robbins:1951}, and
$I_{F}\left(\lambda\right)=  {\mathbb{C}\mathrm{ov}} \left(\nabla_{\lambda}\log q_{\bm \lambda}\left(\bm \theta\right)\right)$.
\end{enumerate}
\end{enumerate}
\end{algorithm}
The performance of VBIL
depends mainly on the variance of the noisy gradient estimator. Following \citet{Tran:2015},
we employ a range of methods, such as control variates and factorisation,
to reduce this variance.

The VB approximation density $q_{\bm \lambda}(\theta) $ for the Archimedean copulas in our article is the inverse gamma discussed in section~\ref{SS: VBIL approx posterior density}.

\section{Simulation studies}\label{S:simulation studies}
\subsection{Performance of the PM and VBIL}\label{S:PM and VBIL}
This section studies the performance of the two new approaches PM
and VBIL for estimating high dimensional Clayton and Gumbel copulas in various simulation
settings. Data are generated from both Clayton and Gumbel copulas
with all the discrete margins following a Bernoulli distribution.
Various simulation scenarios are considered: $J\in\left\{ 10,25,50\right\} $
with $n\in\left\{ 250,500,1000\right\} $ and $J=100$ with $n\in\left\{ 250,500\right\} $. The true value of $\theta$ for the Gumbel was 1.25 and for the Clayton it was 1.
The posterior distribution of $\theta$ is estimated using the correlated
and block PM and VBIL methods. Each MCMC chain consisted of 11000 iterates
with the first 1000 iterates used as burnin. We set $G=100$ blocks
and $\phi=0.9999$ for the block and correlated PM, respectively.
\citet{Tran:2016} show that the optimal number of points $M$ is selected
such that the variance of the log of the likelihood estimate $\sigma_{opt}^{2}$
is approximately $2.16^{2}/\left(1-\widehat{\rho}^{2}\right)$ where
$\widehat{\rho}$ is the estimated correlation between $\log\widehat{L}_{M}\left(\theta',\bm u'\right)$
and $\log\widehat{L}_{M}\left(\theta,\bm u\right)$. 
We used the adaptive random walk method with automatic scaling of \citet{Garthwaite:2015}
to ensure that the overall acceptance rate was around 0.44 \citep{Roberts:1997}. For VBIL,
we used the inverse gamma distribution for $q_{\boldsymbol{\lambda}}\left(\theta\right)$
and we set $S=140$, the learning rate $a_{t}=\frac{1}{10+t}$, where
$t$ is the iteration number, and fix the number of VBIL iterations
to $50$. In this example, the parameters of the marginal distributions
are set to their true values. The PM and VBIL methods are implemented in Matlab and are run on 28 CPU-cores of a high performance computer cluster.

To define our measure of the inefficiency of a PM scheme that takes into account
the computing time, we first define the Integrated Autocorrelation
Time (IACT). For a univariate parameter $\theta$, the IACT is defined
as 
\begin{equation}
\textrm{IACT}_{\theta}=1+2\sum_{t=1}^{\infty}\rho_{\theta}\left(t\right),
\end{equation}
where $\rho_{\theta}\left(t\right)$ are the autocorrelations of the iterates
of $\theta$ in the MCMC after the chain has converged. A large value
of IACT indicates that the chain does not mix well. We estimate $\textrm{IACT}_{\theta}$
based on $R$ iterates of MCMC $\theta^{\left[1\right]},...,\theta^{\left[R\right]}$
after convergence as 
\begin{equation}
\widehat{\textrm{IACT}}_{\theta}=1+2\sum_{t=1}^{L^{*}}\widehat{\rho}_{\theta}\left(t\right),
\end{equation}
where $\widehat{\rho}_{\theta}\left(t\right)$ is the estimate of
$\rho_{\theta}\left(t\right)$, $L^{*}=\textrm{min}\left(1000,L\right)$
and $L=\textrm{min}_{t\leq R}|\widehat{\rho}_{\theta}\left(t\right)|<2/\sqrt{R}$
because $1/\sqrt{R}$ is approximately the standard error of the autocorrelation
estimates when the series is white noise. Our measure of the inefficiency
of a sampler is the time normalised variance (TNV) defined
as
\begin{equation}
\textrm{TNV}=\widehat{\textrm{IACT}}_{\theta}\times\textrm{CT},
\end{equation}
where CT is the computing time.

\begin{figure}[H]
\caption{Kernel smoothing density estimates of the posterior density of the
Clayton copula parameter $\theta$ for $J=\left\{ 10,25,50\right\} $
dimensions with $n=1000$ observations and $J=100$ dimensions with
$n=500$ observations estimated using block PM, correlated PM and
VBIL methods\label{fig:Kernel-smoothing-density clayton}}

\includegraphics[width=15cm,height=15cm]{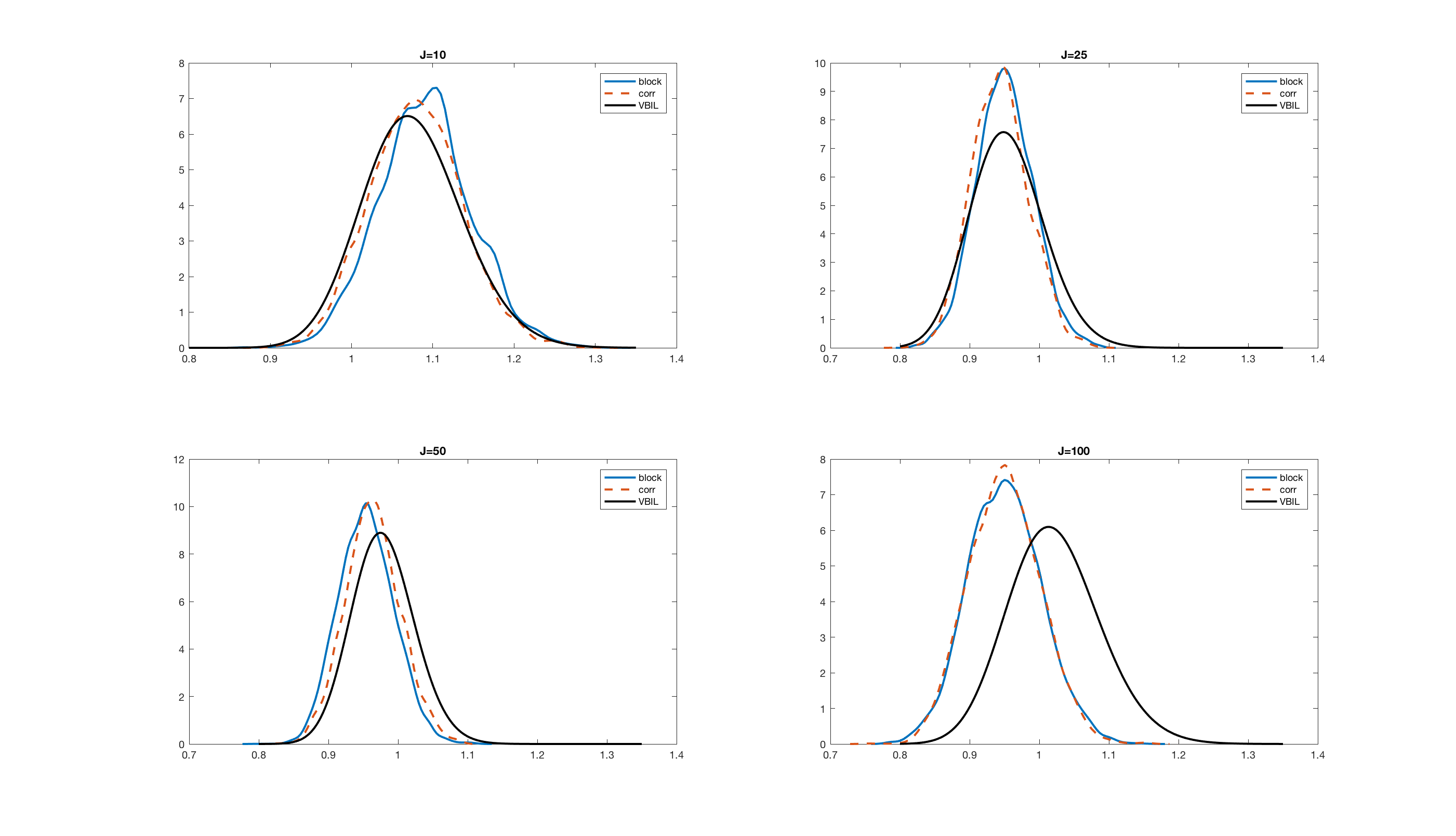}
\end{figure}

\begin{figure}[H]
\caption{Kernel smoothing density estimates of the posterior density of the
Gumbel copula parameter $\theta$ for $J=\left\{ 10,25,50\right\} $
dimensions with $n=1000$ observations and $J=100$ dimensions with
$n=500$ observations estimated using block PM, correlated PM and
VBIL methods\label{fig:Kernel-smoothing-density gumbel}}

\includegraphics[width=15cm,height=15cm]{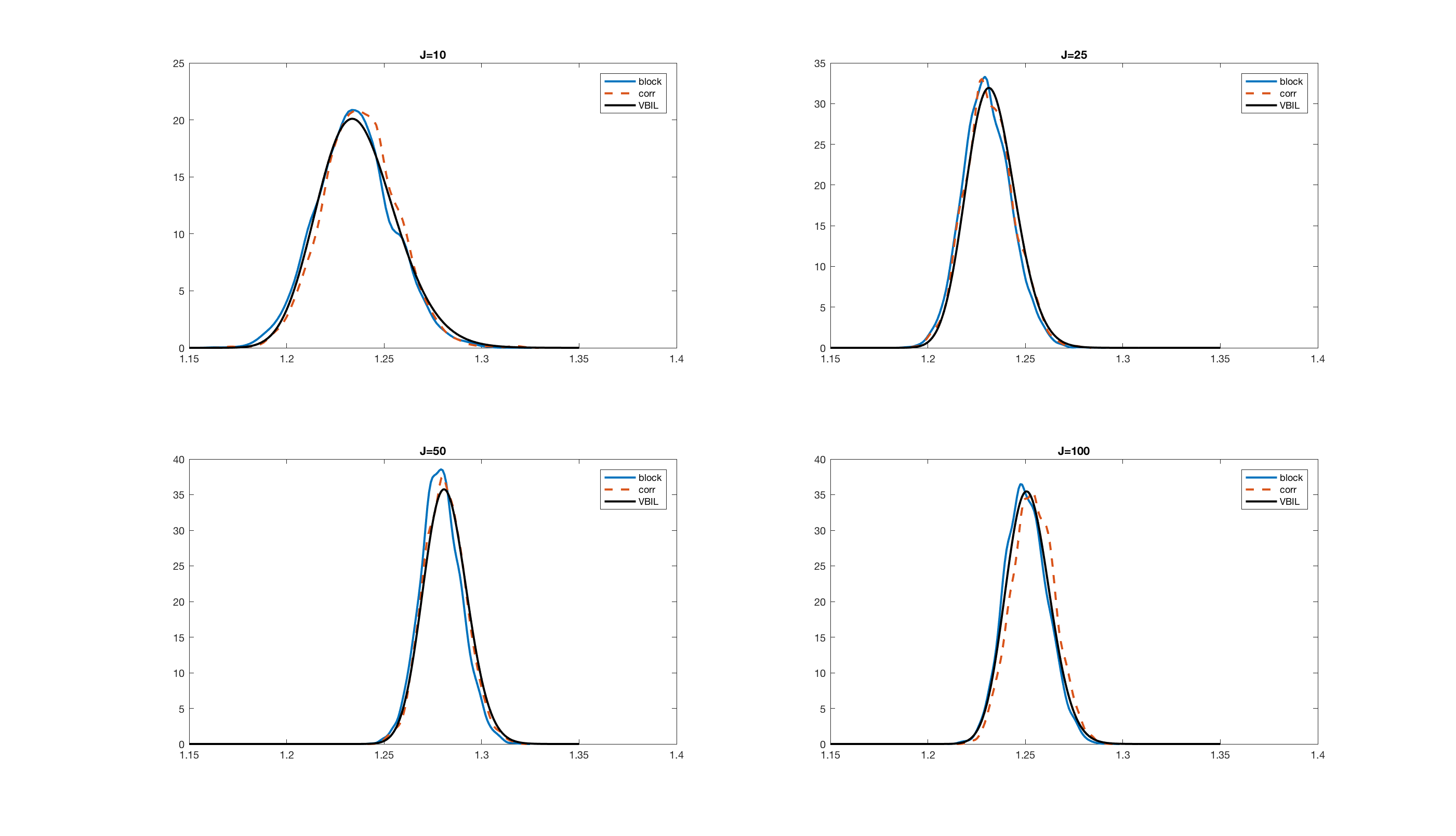}
\end{figure}

Tables \ref{tab:Estimates-of-the 10 dimensions sim1} to \ref{tab:Estimates-of-the 100 dimensions sim1}
summarize the simulation results for $J=\left\{ 10,25,50\right\} $
with $n=\left\{ 250,500,1000\right\} $ and $J=100$ with $n\in\left\{ 250,500\right\} $
for both Gumbel and Clayton copulas. Overall, they show that both
correlated and block PM estimates of the posterior mean of $\theta$
are close to the true values, as are the VBIL estimates. The tables
also show that the IACT's for parameter $\theta$ are small, which
indicates that the chains mixed well. The block PM approach is always better
than the correlated PM in terms of TNV. Block PM takes less CPU time in each MCMC iteration as it
only updates a block of $\boldsymbol{u}_{\left(k\right)}$ whereas
the correlated PM updates the entire set of random numbers $\boldsymbol{u}$.
The VBIL approach is the best in terms of CPU time for all simulation
settings. In this simulation example, we fix the number of iteration
of VBIL to $50$, the CPU time of the VBIL approach can be much lower
as it often converges less than 20 iterations. Figures \ref{fig:Kernel-smoothing-density clayton}
and \ref{fig:Kernel-smoothing-density gumbel} plot some of the estimates
of the posterior marginal densities $\pi\left(\theta\right)$ of $\theta$
for the PM methods and VBIL. The MCMC density estimates are obtained using the Matlab kernel density function \texttt{ksdensity}.
The posterior estimates for both PM methods are very similar. The
VBIL estimates are also very close to the PM estimates, even for $J=50$
and $100$ for the Gumbel copula. However, there is more of a discrepancy
between VBIL estimates and the PM estimates for $J=100$ for the Clayton
copula. 

\begin{table}[H]
\caption{Estimates of the posterior mean (with posterior standard deviation
in parentheses) for a Clayton and Gumbel copula with $J=10$ with $n=\left\{ 250,500,1000\right\} $.
The $\textrm{rel}\;\textrm{TNV}=\textrm{TNV}_{\textrm{method}}/\textrm{TNV}_{\textrm{block}}$.
The CPU time is in minutes.\label{tab:Estimates-of-the 10 dimensions sim1}}

\centering{}%
\begin{tabular}{c|c|ccc|ccc}
\hline 
 &  & \multicolumn{3}{c|}{Clayton} & \multicolumn{3}{c}{Gumbel}\tabularnewline
\hline 
$n$ &  & Corr & Block & VBIL & Corr & Block & VBIL\tabularnewline
\hline 
$250$ & Est. & $\underset{\left(0.122\right)}{1.175}$ & $\underset{\left(0.117\right)}{1.178}$ & $\underset{\left(0.129\right)}{1.167}$ & $\underset{\left(0.0464\right)}{1.344}$ & $\underset{\left(0.0451\right)}{1.347}$ & $\underset{\left(0.050\right)}{1.343}$\tabularnewline
 & $M$ & $50$ & $50$ & $50$ & $50$ & $50$ & $50$\tabularnewline
 & CPU time & $38.500$ & $34.833$ & $0.425$ & $45.100$ & $38.500$ & $0.478$\tabularnewline
 & IACT & $5.644$ & $4.992$ &  & $7.550$ & $6.034$ & \tabularnewline
 & TNV & $217.294$ & $173.886$ &  & $340.505$ & $232.309$ & \tabularnewline
 & Rel. TNV & $1.250$ & $1$ &  & $1.466$ & $1$ & \tabularnewline
\hline 
$500$ & Est. & $\underset{\left(0.073\right)}{0.966}$ & $\underset{\left(0.073\right)}{0.964}$ & $\underset{\left(0.0773\right)}{0.957}$ & $\underset{\left(0.0289\right)}{1.2790}$ & $\underset{\left(0.0287\right)}{1.2777}$ & $\underset{\left(0.032\right)}{1.278}$\tabularnewline
 & $M$ & $50$ & $50$ & $50$ & $50$ & $50$ & $50$\tabularnewline
 & CPU time & $40.133$ & $37.217$ & $0.692$ & $47.666$ & $40.333$ & $0.977$\tabularnewline
 & IACT & $5.020$ & $4.555$ &  & $8.425$ & $6.320$ & \tabularnewline
 & TNV & $201.468$ & $169.523$ &  & $401.586$ & $254.905$ & \tabularnewline
 & Rel. TNV & $1.188$ & $1$ &  & $1.575$ & $1$ & \tabularnewline
\hline 
$1000$ & Est. & $\underset{\left(0.057\right)}{1.080}$ & $\underset{\left(0.057\right)}{1.091}$ & $\underset{\left(0.062\right)}{1.077}$ & $\underset{\left(0.0188\right)}{1.2378}$ & $\underset{\left(0.0196\right)}{1.2349}$ & $\underset{\left(0.020\right)}{1.237}$\tabularnewline
 & $M$ & $50$ & $50$ & $50$ & $50$ & $50$ & $50$\tabularnewline
 & CPU time & $42.167$ & $38.500$ & $1.238$ & $49.500$ & $40.333$ & $1.805$\tabularnewline
 & IACT & $7.094$ & $5.351$ &  & $5.709$ & $5.293$ & \tabularnewline
 & TNV & $299.133$ & $206.014$ &  & $282.596$ & $213.483$ & \tabularnewline
 & Rel. TNV & $1.452$ & $1$ &  & $1.324$ & $1$ & \tabularnewline
\hline 
\end{tabular}
\end{table}

\begin{table}[H]
\caption{Estimates of the posterior mean (with posterior standard deviation
in parentheses) for a Clayton and Gumbel copula with $J=25$ with $n=\left\{ 250,500,1000\right\} $.
The $\textrm{rel}\;\textrm{TNV}=\textrm{TNV}_{\textrm{method}}/\textrm{TNV}_{\textrm{block}}$.
The CPU time is in minutes. \label{tab:Estimates-of-the 25 dimensions sim1}}

\centering{}%
\begin{tabular}{c|c|ccc|ccc}
\hline 
 &  & \multicolumn{3}{c|}{Clayton} & \multicolumn{3}{c}{Gumbel}\tabularnewline
\hline 
$n$ &  & Corr & Block & VBIL & Corr & Block & VBIL\tabularnewline
\hline 
$250$ & Est. & $\underset{\left(0.091\right)}{1.077}$ & $\underset{\left(0.092\right)}{1.065}$ & $\underset{\left(0.094\right)}{1.072}$ & $\underset{\left(0.022\right)}{1.337}$ & $\underset{\left(0.021\right)}{1.333}$ & $\underset{\left(0.025\right)}{1.333}$\tabularnewline
 & $M$ & $250$ & $250$ & $250$ & $250$ & $250$ & $250$\tabularnewline
 & CPU time & $49.500$ & $38.702$ & $1.453$ & $34.833$ & $20.333$ & $1.805$\tabularnewline
 & IACT & $7.916$ & $7.209$ &  & $5.569$ & $7.578$ & \tabularnewline
 & TNV & $391.842$ & $279.003$ &  & $193.985$ & $154.083$ & \tabularnewline
 & Rel. TNV & $1.404$ & $1$ &  & $1.259$ & $1$ & \tabularnewline
\hline 
$500$ & Est. & $\underset{\left(0.062\right)}{0.959}$ & $\underset{\left(0.058\right)}{0.957}$ & $\underset{\left(0.065\right)}{0.948}$ & $\underset{\left(0.019\right)}{1.272}$ & $\underset{\left(0.019\right)}{1.274}$ & $\underset{\left(0.021\right)}{1.278}$\tabularnewline
 & $M$ & $250$ & $250$ & $250$ & $250$ & $250$ & $250$\tabularnewline
 & CPU time & $62.333$ & $40.165$ & $2.741$ & $69.667$ & $42.442$ & $3.929$\tabularnewline
 & IACT & $6.815$ & $4.713$ &  & $6.801$ & $6.204$ & \tabularnewline
 & TNV & $424.799$ & $189.298$ &  & $473.805$ & $263.310$ & \tabularnewline
 & Rel. TNV & $2.244$ & $1$ &  & $1.799$ & $1$ & \tabularnewline
\hline 
$1000$ & Est. & $\underset{\left(0.041\right)}{0.944}$ & $\underset{\left(0.041\right)}{0.951}$ & $\underset{\left(0.053\right)}{0.954}$ & $\underset{\left(0.013\right)}{1.2317}$ & $\underset{\left(0.012\right)}{1.2303}$ & $\underset{\left(0.013\right)}{1.233}$\tabularnewline
 & $M$ & $250$ & $250$ & $250$ & $250$ & $250$ & $250$\tabularnewline
 & CPU time & $91.666$ & $42.167$ & $5.315$ & $102.667$ & $47.666$ & $4.943$\tabularnewline
 & IACT & $4.618$ & $5.576$ &  & $5.131$ & $6.656$ & \tabularnewline
 & TNV & $423.313$ & $235.123$ &  & $526.784$ & $317.265$ & \tabularnewline
 & Rel. TNV & $1.800$ & $1$ &  & $1.660$ & $1$ & \tabularnewline
\hline 
\end{tabular}
\end{table}

\begin{table}[H]
\caption{Estimates of the posterior mean (with posterior standard deviation
in parentheses) for a Clayton and Gumbel copula with $J=50$ with $n=\left\{ 250,500,1000\right\} $.
The $\textrm{rel}\;\textrm{TNV}=\textrm{TNV}_{\textrm{method}}/\textrm{TNV}_{\textrm{block}}$.
The CPU time is in minutes.\label{tab:Estimates-of-the 50 dimensions sim1}}

\centering{}%
\begin{tabular}{c|c|ccc|ccc}
\hline 
 &  & \multicolumn{3}{c|}{Clayton} & \multicolumn{3}{c}{Gumbel}\tabularnewline
\hline 
$n$ &  & Corr & Block & VBIL & Corr & Block & VBIL\tabularnewline
\hline 
$250$ & Est. & $\underset{\left(0.079\right)}{0.982}$ & $\underset{\left(0.078\right)}{0.981}$ & $\underset{\left(0.086\right)}{1.008}$ & $\underset{\left(0.020\right)}{1.279}$ & $\underset{\left(0.021\right)}{1.276}$ & $\underset{\left(0.022\right)}{1.282}$\tabularnewline
 & $M$ & $500$ & $500$ & $500$ & $500$ & $500$ & $500$\tabularnewline
 & CPU time & $83.783$ & $40.517$ & $5.294$ & $86.167$ & $44.917$ & $3.864$\tabularnewline
 & IACT & $8.850$ & $5.449$ &  & $5.352$ & $4.969$ & \tabularnewline
 & TNV & $741.480$ & $220.777$ &  & $461.166$ & $223.192$ & \tabularnewline
 & Rel. TNV & $3.359$ & $1$ &  & $2.066$ & $1$ & \tabularnewline
\hline 
$500$ & Est. & $\underset{\left(0.064\right)}{1.066}$ & $\underset{\left(0.059\right)}{1.040}$ & $\underset{\left(0.070\right)}{1.079}$ & $\underset{\left(0.0134\right)}{1.2508}$ & $\underset{\left(0.0134\right)}{1.247}$ & $\underset{\left(0.016\right)}{1.252}$\tabularnewline
 & $M$ & $500$ & $500$ & $500$ & $500$ & $500$ & $500$\tabularnewline
 & CPU time & $133.833$ & $45.833$ & $10.391$ & $133.833$ & $47.666$ & $12.345$\tabularnewline
 & IACT & $7.120$ & $6.671$ &  & $5.143$ & $6.365$ & \tabularnewline
 & TNV & $952.891$ & $305.752$ &  & $688.303$ & $303.394$ & \tabularnewline
 & Rel. TNV & $3.117$ & $1$ &  & $2.269$ & $1$ & \tabularnewline
\hline 
$1000$ & Est. & $\underset{\left(0.040\right)}{0.964}$ & $\underset{\left(0.040\right)}{0.955}$ & $\underset{\left(0.045\right)}{0.978}$ & $\underset{\left(0.011\right)}{1.281}$ & $\underset{\left(0.010\right)}{1.279}$ & $\underset{\left(0.011\right)}{1.281}$\tabularnewline
 & $M$ & $500$ & $500$ & $500$ & $500$ & $500$ & $500$\tabularnewline
 & CPU time & $247.867$ & $62.333$ & $18.794$ & $238.333$ & $78.933$ & $24.944$\tabularnewline
 & IACT & $5.507$ & $5.859$ &  & $5.347$ & $6.927$ & \tabularnewline
 & TNV & $1365.004$ & $365.209$ &  & $1274.367$ & $477.499$ & \tabularnewline
 & Rel. TNV & $3.738$ & $1$ &  & $2.668$ & $1$ & \tabularnewline
\hline 
\end{tabular}
\end{table}

\begin{table}[H]
\caption{Estimates of the posterior mean (with posterior standard deviation
in parentheses) for a Clayton and Gumbel copula with $J=100$ with $n=\left\{ 250,500\right\} $.
The $\textrm{rel}\;\textrm{TNV}=\textrm{TNV}_{\textrm{method}}/\textrm{TNV}_{\textrm{block}}$.
The CPU time is the time in minutes. \label{tab:Estimates-of-the 100 dimensions sim1}}

\centering{}%
\begin{tabular}{c|c|ccc|ccc}
\hline 
\multicolumn{1}{c}{} &  & \multicolumn{3}{c|}{Clayton} & \multicolumn{3}{c}{Gumbel}\tabularnewline
\hline 
$n$ &  & Corr & Block & VBIL & Corr & Block & VBIL\tabularnewline
\hline 
$250$ & Est. & $\underset{\left(0.083\right)}{1.055}$ & $\underset{\left(0.073\right)}{1.010}$ & $\underset{\left(0.0854\right)}{1.083}$ & $\underset{\left(0.017\right)}{1.313}$ & $\underset{\left(0.018\right)}{1.311}$ & $\underset{\left(0.019\right)}{1.314}$\tabularnewline
 & $M$ & $1000$ & $1000$ & $1000$ & $2500$ & $2500$ & $2500$\tabularnewline
 & CPU time & $207.167$ & $62.883$ & $19.876$ & $524.333$ & $187.583$ & $70.054$\tabularnewline
 & IACT & $11.500$ & $5.910$ &  & $4.751$ & $7.128$ & \tabularnewline
 & TNV & $2382.421$ & $371.639$ &  & $2491.106$ & $1337.972$ & \tabularnewline
 & Rel. TNV & $6.411$ & $1$ &  & $1.862$ & $1$ & \tabularnewline
\hline 
$500$ & Est. & $\underset{\left(0.051\right)}{0.950}$ & $\underset{\left(0.053\right)}{0.951}$ & $\underset{\left(0.066\right)}{1.020}$ & $\underset{\left(0.011\right)}{1.254}$ & $\underset{\left(0.011\right)}{1.248}$ & $\underset{\left(0.011\right)}{1.252}$\tabularnewline
 & $M$ & $2500$ & $2500$ & $1000$ & $2500$ & $2500$ & $2500$\tabularnewline
 & CPU time & $968$ & $229.71$ & $34.471$ & $957$ & $288.966$ & $125.391$\tabularnewline
 & IACT & $6.364$ & $8.272$ &  & $4.922$ & $8.359$ & \tabularnewline
 & TNV & $6160.352$ & $2129.871$ &  & $4710.354$ & $2415.467$ & \tabularnewline
 & Rel. TNV & $2.892$ & $1$ &  & $1.950$ & $1$ & \tabularnewline
\hline 
\end{tabular}
\end{table}

\subsection{Comparison of the PM and data augmentation}\label{S:PM and VBIL vs data augmentation}
This section compares the pseudo marginal and data augmentation approaches.
\cite{Pitt2006} proposed Algorithm~\ref{alg: data augmentation}  as an efficient Bayesian data augmentation (DA) method to estimate the
 parameters of a Gaussian copula with discrete marginals and \cite{Smith2012} generalized it to other copulas.
 Section~\ref{S: DA approach supplement} gives more details of the algorithm.
 \begin{algorithm} \caption{Data Augmentation}\label{alg: data augmentation}
\begin{itemize}
\item
Generate the $j$ marginal $\bm u_{(j)}, j=1, \dots, J$, from
$p\left(\boldsymbol{u}_{\left(j\right)}|\boldsymbol{\theta},\boldsymbol{u}_{\left(k\neq j\right)},\boldsymbol{x}\right)$
for $j=1, \dots, J$.

\item
Generate $\bm \theta$ from
$p\left(\boldsymbol{\theta}|\boldsymbol{u}\right)$
\end{itemize}
\end{algorithm}

We now compare the block PM method to the data augmentation method.
Data are generated from both Clayton and Gumbel copulas with all the
discrete margins following a Bernoulli distribution. Various simulation
scenarios are considered: $J\in\left\{ 5,10,15\right\} $ with $n\in\left\{ 250,500\right\} $.
Each of the MCMC chain consisted of 11000 iterates with the first
1000 iterates used as burnin. The parameters of the marginal distributions
were set to their true values. 

Tables \ref{tab:Estimates-of-the dataaugsim5} to \ref{tab:Estimates-of-the dataaugsim15}
summarise the simulation results and show that (i) The estimates
from the two methods are close to each other; (ii) the PM method is
much faster than the data augmentation method; (iii) the PM method
has a much smaller IACT value. The data augmentation approach generates
the parameter $\theta$ conditioned on the latent variables $\boldsymbol{u}$
in a Metropolis within Gibbs step. As shown, this is very inefficient because the
latent variables $\boldsymbol{u}$ is highly correlated with the parameter
$\theta$. The PM method is much more efficient because it generates
the parameter $\theta$ by integrating out the latent variables $\boldsymbol{u}$;
(iv) the time normalised variance of the PM method is much smaller
than that of the data augmentation method. 

\begin{table}[H]
\caption{Estimates of the posterior mean (with posterior standard deviation
in parentheses) for a Clayton and Gumbel copula with $J=5$ with $n=\left\{ 250,500\right\} $.
The $\textrm{rel}\;\textrm{TNV}=\textrm{TNV}_{\textrm{method}}/\textrm{TNV}_{\textrm{block}}$.
The CPU time is the time in minutes.\label{tab:Estimates-of-the dataaugsim5}}

\centering{}%
\begin{tabular}{c|c|cc|cc}
\hline 
 &  & \multicolumn{2}{c|}{Clayton} & \multicolumn{2}{c}{Gumbel}\tabularnewline
\hline 
$n$ &  & Data Aug. & Block PM & Data Aug. & Block PM\tabularnewline
\hline 
$250$ & Est. & $\underset{\left(0.121\right)}{0.922}$ & $\underset{\left(0.113\right)}{0.837}$ & $\underset{\left(0.053\right)}{1.236}$ & $\underset{\left(0.052\right)}{1.226}$\tabularnewline
 & $M$ & - & $50$ & - & $50$\tabularnewline
 & CPU time & $185.166$ & $32.083$ & $221.833$ & $34.283$\tabularnewline
 & IACT & $102.606$ & $4.764$ & $21.239$ & $4.872$\tabularnewline
 & TNV & $18999.143$ & $152.843$ & $4711.511$ & $167.027$\tabularnewline
 & Rel. TNV & $124.304$ & $1$ & $28.208$ & $1$\tabularnewline
\hline 
$500$ & Est. & $\underset{\left(0.101\right)}{1.116}$ & $\underset{\left(0.104\right)}{1.130}$ & $\underset{\left(0.040\right)}{1.272}$ & $\underset{\left(0.035\right)}{1.217}$\tabularnewline
 & $M$ & - & $50$ & - & $50$\tabularnewline
 & CPU time & $201.667$ & $34.484$ & $245.667$ & $35.750$\tabularnewline
 & IACT & $69.823$ & $5.230$ & $20.680$ & $5.479$\tabularnewline
 & TNV & $14080.995$ & $180.351$ & $5080.394$ & $195.874$\tabularnewline
 & Rel. TNV & $78.076$ & $1$ & $25.937$ & $1$\tabularnewline
\hline 
\end{tabular}
\end{table}

\begin{table}[H]
\caption{Estimates of the posterior mean (with posterior standard deviation
in parentheses) for a Clayton and Gumbel copula with $J=10$ with $n=\left\{ 250,500\right\} $.
The $\textrm{rel}\;\textrm{TNV}=\textrm{TNV}_{\textrm{method}}/\textrm{TNV}_{\textrm{block}}$.
The CPU time is the time in minutes.\label{tab:Estimates-of-the dataaugsim10}}

\centering{}%
\begin{tabular}{c|c|cc|cc}
\hline 
 &  & \multicolumn{2}{c|}{Clayton} & \multicolumn{2}{c}{Gumbel}\tabularnewline
\hline 
$n$ &  & Data Aug. & Block PM& Data Aug. & Block PM\tabularnewline
\hline 
$250$ & Est. & $\underset{\left(0.109\right)}{1.092}$ & $\underset{\left(0.116\right)}{1.171}$ & $\underset{\left(0.039\right)}{1.263}$ & $\underset{\left(0.038\right)}{1.268}$\tabularnewline
 & $M$ & - & $50$ & - & $50$\tabularnewline
 & CPU time & $385.183$ & $32.251$ & $517.000$ & $34.833$\tabularnewline
 & IACT & $85.556$ & $4.670$ & $24.628$ & $5.826$\tabularnewline
 & TNV & $32954.717$ & $150.613$ & $12732.676$ & $202.937$\tabularnewline
 & Rel. TNV & $218.804$ & $1$ & $62.742$ & $1$\tabularnewline
\hline 
$500$ & Est. & $\underset{\left(0.068\right)}{0.959}$ & $\underset{\left(0.070\right)}{0.955}$ & $\underset{\left(0.032\right)}{1.298}$ & $\underset{\left(0.029\right)}{1.270}$\tabularnewline
 & $M$ & - & $50$ & - & $50$\tabularnewline
 & CPU time & $419.833$ & $34.651$ & $806.300$ & $35.200$\tabularnewline
 & IACT & $51.779$ & $5.063$ & $35.457$ & $4.690$\tabularnewline
 & TNV & $21738.533$ & $175.438$ & $28588.979$ & $165.088$\tabularnewline
 & Rel. TNV & $123.910$ & $1$ & $173.174$ & $1$\tabularnewline
\hline 
\end{tabular}
\end{table}

\begin{table}[H]
\caption{Estimates of the posterior mean (with posterior standard deviation
in parentheses) for a Clayton and Gumbel copula with $J=15$ with $n=\left\{ 250,500\right\} $.
The $\textrm{rel}\;\textrm{TNV}=\textrm{TNV}_{\textrm{method}}/\textrm{TNV}_{\textrm{block}}$.
The CPU time is the time in minutes.\label{tab:Estimates-of-the dataaugsim15}}

\centering{}%
\begin{tabular}{c|c|cc|cc}
\hline 
 &  & \multicolumn{2}{c|}{Clayton} & \multicolumn{2}{c}{Gumbel}\tabularnewline
\hline 
$n$ &  & Data Aug. & Block PM& Data Aug. & Block PM\tabularnewline
\hline 
$250$ & Est. & $\underset{\left(0.095\right)}{1.134}$ & $\underset{\left(0.098\right)}{1.074}$ & $\underset{\left(0.038\right)}{1.328}$ & $\underset{\left(0.032\right)}{1.298}$\tabularnewline
 & $M$ & - & $50$ & $-$ & $50$\tabularnewline
 & CPU time & $598.217$ & $33.251$ & $1079.833$ & $35.254$\tabularnewline
 & IACT & $98.925$ & $5.612$ & $37.317$ & $5.299$\tabularnewline
 & TNV & $59178.616$ & $186.605$ & $40296.128$ & $186.811$\tabularnewline
 & Rel. TNV & $317.133$ & $1$ & $215.705$ & $1$\tabularnewline
\hline 
$500$ & Est. & $\underset{\left(0.068\right)}{1.012}$ & $\underset{\left(0.067\right)}{1.014}$ & $\underset{\left(0.024\right)}{1.283}$ & $\underset{\left(0.023\right)}{1.275}$\tabularnewline
 & $M$ & - & $50$ & - & $50$\tabularnewline
 & CPU time & $652.667$ & $35.201$ & $1413.500$ & $35.456$\tabularnewline
 & IACT & $69.031$ & $5.906$ & $28.946$ & $7.276$\tabularnewline
 & TNV & $45054.256$ & $207.897$ & $40915.171$ & $257.978$\tabularnewline
 & Rel. TNV & $216.714$ & $1$ & $158.599$ & $1$\tabularnewline
\hline 
\end{tabular}
\end{table}

\section{Real-data examples}\label{S:applications}
\subsection{HILDA data}\label{SS: modeling well being data}
The data used in the examples is obtained from the Household, Income,
and Labour Dynamics in Australia (HILDA) survey for the year 2013. We use 50 categorical variables
which include a range of well-being attributes, such as health (items
1-36), income (item 50), education (item 49), and life satisfaction
(item 37), and a range of major life-shock events, such as getting
married (item 38), separated from spouse (item 39), getting back together
with the spouse (item 40), serious personal injury (item 41), death
of spouse or child (item 42), death of a close friend (item 43), being
a victim of a property crime (item 44), promotion at work (item 45),
major improvement (item 46) and major worsening in personal finances
(item 47), and change of residence (item 48). We transformed the response
of each person to each item into a 0 or 1. Thus, for questions on
health we classified a person as healthy (0) or unhealthy (1). Similarly,
for income (item 50), education (item 49), and life satisfaction variables
(item 37), we classified people into rich (0) or poor (1), high education
(0) or low education (1), and high life satisfaction (0) or low
life satisfaction (1). In this example, the unit of analysis was a
male aged above 45, who has non-missing information on all the variables
being considered, and who is not in the labour force and married,
resulting in $n=1210$ individuals.
Section~\ref{S: data set on well being} gives further details on this dataset.

\subsubsection{Discrete Clayton and Gumbel copulas }  \label{SSS: clayton analysis}
We estimated the joint binary distribution of the well-being attributes
and life shock events by fitting Clayton and Gumbel copula models using
the correlated and blocked PM
methods and the VBIL method. In this example, the parameters of the marginal distributions were set to their sample estimates.
Each MCMC chain consisted of 11,000 iterates
with the first 1000 iterates used as burnin.  
For VBIL, the variational distribution $q_{\bm \lambda}\left( \theta\right)$ was
the inverse gamma distribution $IG\left(a,b\right)$, with $S=140$ samples used
to estimate the gradient of the lower bound. We fix the number of VBIL iterations to 25.

Table~\ref{tab:The-variance-of clayton well-being} shows the variance of the log of the estimated likelihood for
different numbers of samples $M$ for the 50 dimensional Clayton and Gumbel copulas. In particular, the table shows
that even with $M=16384$ standard PM would get stuck. We therefore do not report results
for the standard PM method in this section and the next as their TNV would be much higher than that of the correlated or block PM methods.
We can not even use the data augmentation method in this example as it is so computational expensive to do so.

Table \ref{tab:Estimates-of-the clayton and gumbel well being}  summarizes the
estimation results and show that:  (i) ~The block PM is better than that of the correlated PM method in terms of the time normalized variance TNV, (ii)~The VBIL method is at least five times as fast as the block PM method. (iii)~All the estimates are close to each other.

\begin{table}[H]
\caption{The variance of log of the estimated likelihood, evaluated at a posterior mean estimate, for the discrete Clayton copula, discrete Gumbel copula and the mixed marginal Gumbel copula as a function of the number of points $M$. \label{tab:The-variance-of clayton well-being}}
\centering{}%
\begin{tabular}{ccc||cc}
\hline
 $M$& Clayton   &  Gumbel &$M$&Mixed Gumbel\tabularnewline
\hline
256 & 88.75  & 250.71 & 64 & 47.90 \tabularnewline
512  & 51.40  & 151.92  & 128 & 37.70 \tabularnewline
1024  & 41.56  & 131.22 & 256 & 17.03 \tabularnewline
2048  & 20.95   & 106.31 & 512 & 12.36  \tabularnewline
8192  & 11.49  & 52.86 & 1024 & 11.03 \tabularnewline
16384  & 8.51  & 33.77& 16384 & 2.55 \tabularnewline
\hline
\end{tabular}
\end{table}

\begin{table}[H]
\caption{Estimates of the posterior mean (with posterior standard deviation
in parentheses) for the Clayton and Gumbel copula with $J=50$ with $n=1210$.
The $\textrm{rel}\;\textrm{TNV}=\textrm{TNV}_{\textrm{method}}/\textrm{TNV}_{\textrm{block}}$.
The CPU time is in minutes.\label{tab:Estimates-of-the clayton and gumbel well being}}

\centering{}%
\begin{tabular}{c|ccc|ccc}
\hline 
 & \multicolumn{3}{c|}{Clayton} & \multicolumn{3}{c}{Gumbel}\tabularnewline
\hline 
 & Corr & Block & VBIL & Corr & Block & VBIL\tabularnewline
\hline 
Est. & $\underset{\left(0.011\right)}{0.389}$ & $\underset{\left(0.011\right)}{0.395}$ & $\underset{\left(0.012\right)}{0.386}$ & $\underset{\left(0.010\right)}{1.208}$ & $\underset{\left(0.010\right)}{1.204}$ & $\underset{\left(0.008\right)}{1.200}$\tabularnewline
$M$ & $1024$ & $1024$ & $1024$ & $2048$ & $2048$ & $2048$\tabularnewline
CPU time & $471.167$ & $119.166$ & $22.162$ & $1001.520$ & $264.325$ & $37.280$\tabularnewline
IACT & $4.310$ & $5.240$ &  & $20.530$ & $4.930$ & \tabularnewline
TNV & $2030.729$ & $624.430$ &  & $20561.206$ & $1303.202$ & \tabularnewline
Rel. TNV & $3.252$ & $1$ &  & $15.778$ & $1$ & \tabularnewline
\hline 
\end{tabular}
\end{table}

\subsubsection{Mixed marginals}  \label{SS: mixed marginal Gumbel example}
The data used in this example is also obtained from the HILDA survey for the year 2014. We use 50 variables consisting of 30 categorical variables and 20 continuous variables. The continuous variables include income, SF36 continuous health score, weight in kg, height in cm, and hours/mins per week for the following activities: paid employment, travelling to/from paid employment, household errands, housework, outdoor tasks, playing with the children, playing with other people children, volunteer work, and caring for disabled relatives. The categorical variables include community participation activities (11 variables), personal satisfaction variables (8 variables), satisfaction with financial situation, personal safety, employment opportunities, questions about the
current job situation (10 variables), and a question about the availability of internet at home. In this example, we fit a Gumbel copula model for the first $n=1000$ individuals.

Table~\ref{tab:Estimates-of-the mixed real} summarises the
estimation results and shows that: (i)~The block PM sampler is better than the correlated PM in terms of the time normalized variance TNV. (ii)~The VBIL is at least five times faster than the PM approaches. (iii)~All estimates are close to each other.

\begin{table}[H]
\caption{Estimates of the posterior mean (with posterior standard deviation
in parentheses) for a 50 dimensional Gumbel copula model with $n=1000$
for the well-being example with mixed marginals. The $\textrm{rel}\;\textrm{TNV}=\textrm{TNV}_{\textrm{method}}/\textrm{TNV}_{\textrm{block}}$.
The CPU time is the time in minutes. \label{tab:Estimates-of-the mixed real}}

\centering{}%
\begin{tabular}{c|ccc}
\hline 
 & \multicolumn{3}{c}{Gumbel}\tabularnewline
\hline 
 & Corr & Block & VBIL\tabularnewline
\hline 
Est. & $\underset{\left(0.001\right)}{1.012}$ & $\underset{\left(0.001\right)}{1.013}$ & $\underset{\left(0.001\right)}{1.012}$\tabularnewline
$M$ & $128$ & $128$ & $128$\tabularnewline
CPU time & $139.333$ & $97.166$ & $17.314$\tabularnewline
IACT & $4.812$ & $5.987$ & \tabularnewline
TNV & $670.470$ & $581.733$ & \tabularnewline
Rel. TNV & $1.153$ & $1$ & \tabularnewline
\hline 
\end{tabular}
\end{table}

\section{Online supplementary material}
The online supplementary material gives further technical details.

\section{Summary and conclusions}
Our article proposes several computationally efficient methods for estimating
high-dimensional Archimedian copulas, such as Clayton and Gumbel copulas, with discrete or mixed marginals. The proposed methods are based on recent advances in Bayesian computation and work with an unbiased estimate
of the likelihood.
The empirical results suggest that
for a high $nJ$: (a) The PM and VBIL approaches are appreciably more efficient than the data augmentation approach,
which can become computationally infeasible for a large $J$ or $n$;
(b)~The correlated and block PM samplers are much more efficient than the
standard PM sampler; (c)~The block PM sampler always performs better than the correlated PM sampler; (d)~The VBIL method is the fastest method,
and usually produces good approximations of the posterior. 

\bibliographystyle{apalike}
\bibliography{references_v1}

\pagebreak
\renewcommand{\theequation}{S\arabic{equation}}
\renewcommand{\thesection}{S\arabic{section}}
\renewcommand{\theproposition}{S\arabic{proposition}}
\renewcommand{\theassumption}{S\arabic{assumption}}
\renewcommand{\thelemma}{S\arabic{lemma}}
\renewcommand{\thealgorithm}{S\arabic{algorithm}}
\renewcommand{\thefigure}{S\arabic{figure}}
\renewcommand{\thetable}{S\arabic{table}}
\renewcommand{\thepage}{S\arabic{page}}

\setcounter{page}{1}
\setcounter{section}{0}
\setcounter{equation}{0}
\setcounter{algorithm}{0}
\setcounter{table}{0}
\def\myformat#1{\centering#1}

\begin{center}
{\Large \bf Online supplementary material}
\end{center}

All equations, lemmas, tables, etc in the main paper are referred to as equation (1), lemma 1, table 1, etc, and
in this supplement they are referred to as equation (S1), lemma S1 and table S1, etc.

\section{Some further technical results for the Gumbel and Clayton copulas}\label{S: gumbel clayton supplementary material}
\subsection{The Gumbel copula} \label{SS: defining Gumbel copula}
The  $J$-dimensional Gumbel  copula is another popular example of
Archimedean copulas. Its cdf $C\left(\boldsymbol{u}\right) $ and density $c\left(\boldsymbol{u}\right) $ are
\begin{eqnarray*}
C\left(\boldsymbol{u}\right) & := & \exp\left\{ -\left[\sum_{j=1}^{J}\left(-\log\left(u_{j}\right)\right)^{\theta}\right]^{1/\theta}\right\}\\
c\left(\boldsymbol{u}\right) & := & \theta^{J}\exp\left\{ -\left[\sum_{j=1}^{J}\left(-\log\left(u_{j}\right)\right)^{\theta}\right]^{\frac{1}{\theta}}\right\} \\
 & \times & \frac{\prod_{j=1}^{J}\left(-\log\left(u_{j}\right)\right)^{\theta-1}}{\left(\sum_{j=1}^{J}\left(-\log\left(u_{j}\right)\right)^{\theta}\right)^{J}\prod_{j=1}^{J}u_{j}}\times P_{J,\theta}^{G}\left(\left[\sum_{j=1}^{J}\left(-\log\left(u_{j}\right)\right)^{\theta}\right]^{\frac{1}{\theta}}\right),
\end{eqnarray*}
where
\[
P_{J,\theta}^{G}\left(x\right)=\sum_{k=1}^{J}a_{mk}^{G}\left(\theta\right)x^{k},
\]
and
\[
a_{mk}^{G}\left(\theta\right)=\frac{J!}{k!}\sum_{j=1}^{k}\left(\begin{array}{c}
k\\
j
\end{array}\right)\left(\begin{array}{c}
j/\theta\\
J
\end{array}\right)\left(-1\right)^{J-j}.
\]
The dependence parameter $\theta$ is defined on $\left[1,\infty\right)$, where a value of $1$ represents the independence case.
The Gumbel copula is an appropriate choice if the data exhibit weak
correlation at lower values and strong correlation at higher values.

If some of the $a_{j}$ are zero, then directly estimating the integral \eqref{eq:joint prob copula discrete}
is computationally inefficient for the same reasons as given in section~\ref{ss: Clayton copula example}
for the Clayton copula.  It can be readily checked that
\begin{eqnarray*}
D\left(\boldsymbol{u}_{1:K},\boldsymbol{b}_{K+1:J}\right) & := & \theta^{K}\exp\left\{ -\left[\sum_{j=1}^{K}\left(-\log\left(u_{j}\right)\right)^{\theta}+\sum_{j=K+1}^{J}\left(-\log\left(b_{j}\right)\right)^{\theta}\right]^{\frac{1}{\theta}}\right\} \\
 & \times & \frac{\prod_{j=1}^{K}\left(-\log\left(u_{j}\right)\right)^{\theta-1}}{\left(\sum_{j=1}^{K}\left(-\log\left(u_{j}\right)\right)^{\theta}
 +\sum_{j=K+1}^{J}\left(-\log\left(b_{j}\right)\right)^{\theta}\right)^{K}\prod_{j=1}^{K}u_{j}}\\
 & \times & P_{K,\theta}^{G}\left(\left[\sum_{j=1}^{K}\left(-\log\left(u_{j}\right)\right)^{\theta}+\sum_{j=K+1}^{J}\left(-\log\left(b_{j}\right)\right)^{\theta}\right]^{\frac{1}{\theta}}\right).
\end{eqnarray*}
Then, we can rewrite the integral as
\begin{align*}
\int_{a_{1}}^{b_{1}}...\int_{a_{K}}^{b_{K}}D&\left(\boldsymbol{u}_{1:K},\boldsymbol{b}_{K+1:J}\right)  \d\boldsymbol{u}_{1:K}  =\prod_{j=1}^{K}\left(b_{j}-a_{j}\right)\\
\times & \int_{0}^{1}...\int_{0}^{1}D\left(\left(b_{1}-a_{1}\right)v_{1}+a_{1},...,\left(b_{K}-a_{K}\right)v_{K}+a_{K},\boldsymbol{b}_{K+1:J}\right)d\boldsymbol{v}_{1:K}.
\end{align*}
\subsection{The VBIL approximation distribution} \label{SS: VBIL approx posterior density}
For the Clayton copula, the VB approximation to the posterior of $\theta$
is the inverse gamma density
\[
q_{\lambda}\left(\theta\right)=\frac{a^{b}}{\Gamma\left(a\right)}\left(\theta\right)^{-1-a}\exp\left(-b/\theta\right),\;\theta>0,
\]
and for the Gumbel copula
\[
q_{\lambda}\left(\theta\right)=\frac{a^{b}}{\Gamma\left(a\right)}(\theta -1)^{-1-a}\exp\left(-b/\left(\theta-1\right)\right),\;\theta >1,
\]
with the natural parameters $a$ and $b$. The Fisher information
matrix for the inverse gamma is
\begin{align*}
I_{F}\left(a,b\right)=\left(\begin{array}{cc}
\nabla_{aa}\left[\log\Gamma\left(a\right)\right] & -1/b\\
-1/b & a/b^{2}
\end{array}\right)
\end{align*}
with gradient
\begin{align*}
\nabla_{a}\left[\log q_{\lambda}\left(\theta\right)\right]=-\log\left(\theta\right)+\log\left(b\right)-\nabla_{a}\left[\log\Gamma\left(a\right)\right]
\quad \text{and}\quad
\nabla_{b}\left[\log q_{\lambda}\left(\theta\right)\right]=-\frac{1}{\theta}+\frac{a}{b}.
\end{align*}

\section{Further description and analysis of the well-being and life-shock events dataset}  \label{S: data set on well being}
This section gives further details of the of the well-being and life-shock events dataset (abbreviated to \lq well-being dataset\rq)
described in section~\ref{SS: modeling well being data}.
The health data used in this paper is obtained from the SF-36 data
collected by the HILDA survey. The SF-36 (Medical Outcome Trust, Boston,
MA) is a multipurpose and short form health survey with 36 items.
Each item provides multiple choice answers for respondents to select
from in regard to different aspects of their health. SF-36 is one
of the most widely used generic measures of health-related quality
of life (HRQoL) in clinical research and general population health.
It is a standardised questionnaire used to assess patient health across
eight attributes~\citep{Ware1993}. These are physical
functioning (PF, items 3 to 12), role-physical (RP,
items 13 to 16), bodily pain (BP, items 21 and 22), general
health (GH, items 1, 2, 33-36), vitality (VT, items
28-31), social functioning (SF, items 20 and 32), role-emotional
(RE,  items 17 to 19), and mental health (MH, items
23-27). The details of the survey questions can be found in \citet{Ware1993}.

\section{Details of the data augmentation approach} \label{S: DA approach supplement}
This section gives further details of Algorithm~\ref{alg: data augmentation}.
The conditional distribution of $p\left(\boldsymbol{u}_{\left(j\right)}|\boldsymbol{\theta},\boldsymbol{u}_{\left(k\neq j\right)},\boldsymbol{x}\right)$
is given  by
\begin{align*}
p\left(\boldsymbol{u}_{\left(j\right)}|\boldsymbol{\theta},\boldsymbol{u}_{\left(k\neq j\right)},\boldsymbol{x}\right) &  \propto p\left(\boldsymbol{x}|\boldsymbol{\theta},\boldsymbol{u}\right)p\left(\boldsymbol{u}_{\left(j\right)}|\boldsymbol{\theta},\boldsymbol{u}_{\left(k\neq j\right)}\right)\\
 & \propto  \prod_{i=1}^{n}I\left(a_{i,j}\leq u_{i,j}<b_{i,j}\right)c\left(\boldsymbol{u}_{i};\boldsymbol{\theta}\right)\\
 & \propto  \prod_{i=1}^{n}I\left(a_{i,j}\leq u_{i,j}<b_{i,j}\right)c_{j|k\neq j}\left(u_{i,j}|u_{i,k\neq j};\boldsymbol{\theta}\right)
\end{align*}

The latents $u_{i,j}$
are generated from the conditional densities $c_{j|k\neq j}$
constrained to  $\left[a_{i,j},b_{i,j}\right)$
and an iterate of  $\boldsymbol{u}_{\left(j\right)}$
obtained. In this sampling scheme, the copula parameter
$\boldsymbol{\theta}$ is generated conditional on
$\boldsymbol{u}$ from
\begin{align*}
p\left(\boldsymbol{\theta}|\boldsymbol{u},\boldsymbol{x}\right) & =p\left(\boldsymbol{\theta}|\boldsymbol{u}\right)\propto\prod_{i=1}^{n}c\left(\boldsymbol{u}_{i};\boldsymbol{\theta}\right)p\left(\boldsymbol{\theta}\right)
\end{align*}

The following algorithm is  used to generate the latent variables one
 margin at a time.

For  $j=1,...,J$ and for $i=1,...,n$
\begin{itemize}
\item
Compute
 \[A_{ij}=C_{j|\left\{ 1,...,J\right\} \setminus j}\left(a_{i,j}|\left\{ u_{i1},...,u_{iJ}\right\} \setminus u_{ij},\boldsymbol{\theta}\right)\]
and
\[B_{ij}=C_{j|\left\{ 1,...,J\right\} \setminus j}\left(b_{i,j}|\left\{ u_{i1},...,u_{iJ}\right\} \setminus u_{ij},\boldsymbol{\theta}\right)\]
\item
Generate
$w_{i,j}\sim Uniform\left(A_{i,j},B_{i,j}\right)$
\item
Compute
$u_{i,j}=C_{j|\left\{ 1,...,J\right\} \setminus j}^{-1}\left(w_{i,j}|\left\{ u_{i1},...,u_{iJ}\right\} \setminus u_{ij},\boldsymbol{\theta}\right)$
\end{itemize}
\end{document}